\documentclass[10pt, a4paper]{article}
\usepackage{lrec}
\usepackage{multibib}
\newcites{languageresource}{Language Resources}
\usepackage{graphicx}
\usepackage{tabularx}
\usepackage{soul}
\usepackage{titlesec}
\titleformat{\section}{\normalfont\large\bf\center}{\thesection.}{1em}{}
\titleformat{\subsection}{\normalfont\SmallTitleFont\bf\raggedright}{\thesubsection.}{1em}{}
\titleformat{\subsubsection}{\normalfont\normalsize\bf\raggedright}{\thesubsubsection.}{1em}{}
\renewcommand\thesection{\arabic{section}}
\renewcommand\thesubsection{\thesection.\arabic{subsection}}
\renewcommand\thesubsubsection{\thesubsection.\arabic{subsubsection}}

\usepackage{epstopdf}
\usepackage[utf8]{inputenc}

\usepackage{hyperref}
\usepackage{xstring}

\usepackage{color}

\usepackage{amsmath,amssymb}
\usepackage{multirow}

\usepackage[absolute,overlay]{textpos}

\title{Multi-Staged Cross-Lingual Acoustic Model Adaption for Robust Speech Recognition in Real-World Applications --- A Case Study on German Oral History Interviews}

\name{Michael Gref,\textsuperscript{\rm 1,2} Oliver Walter,\textsuperscript{\rm 1} Christoph Schmidt,\textsuperscript{\rm 1} Sven Behnke,\textsuperscript{\rm 1,3} Joachim K{\"o}hler\textsuperscript{\rm 1}}

\address{\textsuperscript{\rm 1}Fraunhofer Institute for Intelligent Analysis and Information Systems (IAIS), Germany\\ 
\textsuperscript{\rm 2}Institute for Pattern Recognition (iPattern), Niederrhein Univ. of Applied Sciences, Germany\\
\textsuperscript{\rm 3}Autonomous Intelligent Systems (AIS), Computer Science Institute VI, Univ. of Bonn, Germany\\
 \{michael.gref, oliver.walter, christoph.andreas.schmidt, sven.behnke, joachim.koehler\}@iais.fraunhofer.de}

\abstract{
While recent automatic speech recognition systems achieve remarkable performance when large amounts of adequate, high quality annotated speech data is used for training, the same systems often only achieve an unsatisfactory result for tasks in domains that greatly deviate from the conditions represented by the training data. For many real-world applications, there is a lack of sufficient data that can be directly used for training robust speech recognition systems. To address this issue, we propose and investigate an approach that performs a robust acoustic model adaption to a target domain in a cross-lingual, multi-staged manner. Our approach enables the exploitation of large-scale training data from other domains in both the same and other languages. We evaluate our approach using the challenging task of German oral history interviews, where we achieve a relative reduction of the word error rate by more than 30\% compared to a model trained from scratch only on the target domain, and 6--7\% relative compared to a model trained robustly on 1000 hours of same-language out-of-domain training data. \\ \newline  }

\begin{document}

\begin{textblock*}{21cm}(0cm,.75cm)
	\begin{center}
	\large In: 12th International Conference on Language Resources and Evaluation (LREC 2020), pages 6354--6362\\
\url{https://www.aclweb.org/anthology/2020.lrec-1.780/}
	\end{center}
\end{textblock*}

\maketitleabstract

\section{Introduction}

Current automatic speech recognition (ASR) systems show remarkable performance on tasks where large amount of annotated and time-aligned speech data is used for training. Other tasks, however, where only little or no data is present---such as under-resourced or zero resource languages---are still a challenge for modern systems. Nevertheless, also within one language, the recognition performance drastically degrades if off-the-shelf systems are applied to different domains in terms of recording conditions and speech characteristics, such as noisy and reverberated recordings, spontaneous, fast speech, accents, unclear pronunciations and age- and health related changes in the ways of speaking---to name but a few. In this work, we propose and investigate an approach that performs a robust acoustic model adaption to a target domain in a cross-lingual, multi-staged manner. The proposed approach utilizes large-scale training data from rich-resourced domains in both the same and other languages.

A lot of investigations in the field of speech recognition are based on well-known, publicly available benchmark data sets like Switchboard, Wall Street Journal and AMI to evaluate proposed systems and approaches. Usually, such data sets cover well-defined, restricted challenges and are of good use to study this specific challenge. However, they are only of limited use for replicating the conditions of challenging real-world applications where many of the aforementioned issues might be present at the same time with varying degrees.

German \textit{oral history} interviews are such a challenging real-world application for automatic speech recognition in which the aforementioned issues are present to an unknown extent and with unknown distribution in each sample. The term oral history, in historical research, refers to conducting and analyzing interviews with contemporary witnesses.

Particularly in Germany, this kind of research focuses above all on the period of the Second World War and National Socialism. But in the meantime, it has also come to include many other topics and historical periods. 
The past forty years have seen a multitude of witnesses to a wide range of historical events interviewed by researchers. In the original sense oral history interviews are characterized by the fact that rather than structuring the interview around questions, the interviewer encourages the interviewee to freely narrate his or her life story. In terms of biographical research, the outcome is qualified as a narrative life-story interview lasting very often at least three or four hours.

For both, archiving and analyzing, time-aligned transcription and indexing with keywords of such interviews is essential. Due to the poor performance of off-shelf recognition systems on such interviews and the lack of sufficient in-domain training data, transcribing, labeling and annotating speech recordings is still often performed manually. The huge effort in terms of time and human resources required to do this, severely limits the efficiency to exploit oral history interviews for digital humanities research. 

As our contribution for solving this problem we developed and studied an approach on this unconstrained, demanding use-case. Further, we investigate the performance of models trained with the proposed approach using German test sets from several other domains. In this way we demonstrate the generalization and robustness of the model for similar related challenges in other domains and the applicability of the system in real applications. In addition, our ablation study experiments show that the cross-lingual adaption of state-of-the-art LF-MMI acoustic models results in significantly improved speech recognition performance---even if only a very small amount of training data from the target domain is used. This observation can contribute to the ongoing research on speech recognition for under-resourced language.

\section{Related Work}

The use of automatic speech recognition technology to transcribe and index oral history interviews has started with the MALACH project \cite{Psutka.2002.MalachProject} where interviews of the \textit{Survivors of the Shoah Visual History Foundation} (VHF) were processed with a state-of-the-art speech recognition in 2002. 
\newcite{Oard.2012.AsrInOralHistory} investigated how speech recognition technology can be used for oral history research. In our prior work \cite{Gref.2018.oralhistory01} we studied the application and adaption of state-of-the-art speech recognition systems to German oral history interviews. Furthermore in that work we proposed an ASR test set for German oral history interviews that we utilize in this work for evaluation. Just recently, \newcite{Pincheny.2019.MalachASRCorpus} proposed parts of data from the MALACH project as a speech recognition challenge for English oral history. The authors agree with our statements, that the challenges of such interviews are still open problems for modern speech recognition systems. In contrast to the German oral history challenge, the MALACH corpus provides a few hundred hours of annotated in-domain training data. For German oral history, such amounts of data are currently not available. Furthermore, in contrast to the German oral history evaluation set used in our work, the interview in the MALACH project are recorded in relatively high-quality audio quality. The German oral history interviews in our evaluation set, however, were conducted by historians throughout many decades with a wide range of recording equipment, setups, and acoustic conditions.

ASR is a popular and highly researched area and new approaches are regularly proposed. End-to-end approaches to speech recognition have received a lot of attention lately. Since these models do not require alignments from a bootstrap model---as conventional HMM-DNN approaches do---these models are significantly simpler to train. In many scenarios, however, conventional HMM-DNN systems perform better than end-to-end systems. This is especially the case when only a small amount of training data is used. 
In ASR, lattice-free maximum mutual information (LF-MMI) trained models \cite{Povey.2016.LF_MMI} achieved state-of-the-art results on many different speech recognition tasks in recent years and are applied in many recently proposed speech recognition systems.    

A common approach for certain out-of-domain tasks is to apply data augmentation on rather clean data and perform multi-condition training. 
\newcite{Ko.2017.KaldiReverbDataAugmentation} studied data augmentation of reverberant speech for state-of-the-art LF-MMI models. Speed perturbation techniques to increase training data variance have been investigated by \newcite{Ko.2015.SpeedPerturbation}. The proposed method in this work is to increase the data three-fold by creating two additional versions of each signal using the constant speed factors $ 0.9 $ and $ 1.1 $---a method used in many recent training routines by default. In \cite{Gref.2018.oralhistory02} we studied noise and reverberation data augmentation for robust speech recognition on German oral history interviews for conventionally (cross-entropy) trained and LF-MMI models.

While data augmentation is quite successful to overcome a mismatch in acoustic conditions between desired applications and training data, it is limited to acoustic distortions that can be artificially created. 
In recent years, transfer learning for acoustic model adaption has raised attention for under-resourced language tasks. 
Transfer learning is an approach to improve generalization and performance by transferring knowledge of a model trained in one scenario to train a model in another related scenario \cite{Goodfellow.2016.DeepLearningBook}. 
It is particularly useful in scenarios where only little training data is available for the main task but a large amount of data is available for a similar or related task. 

\newcite{Wang.2015.transfer_learning_overview_in_speechprocessing} give a detailed overview of transfer learning in speech and language processing. \newcite{Ghahremani.2017.KaldiTransferlearning_LF_MMI} investigated transfer learning using LF-MMI models for many different well-known English speech recognition tasks. For the German oral history task,  we studied transfer learning using a very small amount of training data and a leave-one-speaker-out evaluation approach \cite{Gref.2019.ICME_twostaged_am_adaption}.

Cross-lingual acoustic model adaption or knowledge transfer aims at utilizing the knowledge of models trained in one or more languages to improve speech recognition performance for a target language. \newcite{Xu.2016.semisupervisedCrosslingual} studied semi-supervised learning and cross-lingual knowledge transfer with multilingual data and neural network fine-tuning. \newcite{Feng.2018.improvingCrossLingualKnowledgeTransferability} investigated cross-lingual knowledge transfer in a multilingual setup with language-dependent pre-final layers under each softmax output layer. \newcite{Ma.2017.STTMultiLingualKnowledgeTransfer} have studied multilingual training using LF-MMI models with a joint output layer across languages followed by the adaption to a low-resourced target language.

\section{Proposed Approach}

\begin{figure*}[h]
	\centering
	\includegraphics[width=0.9\textwidth]{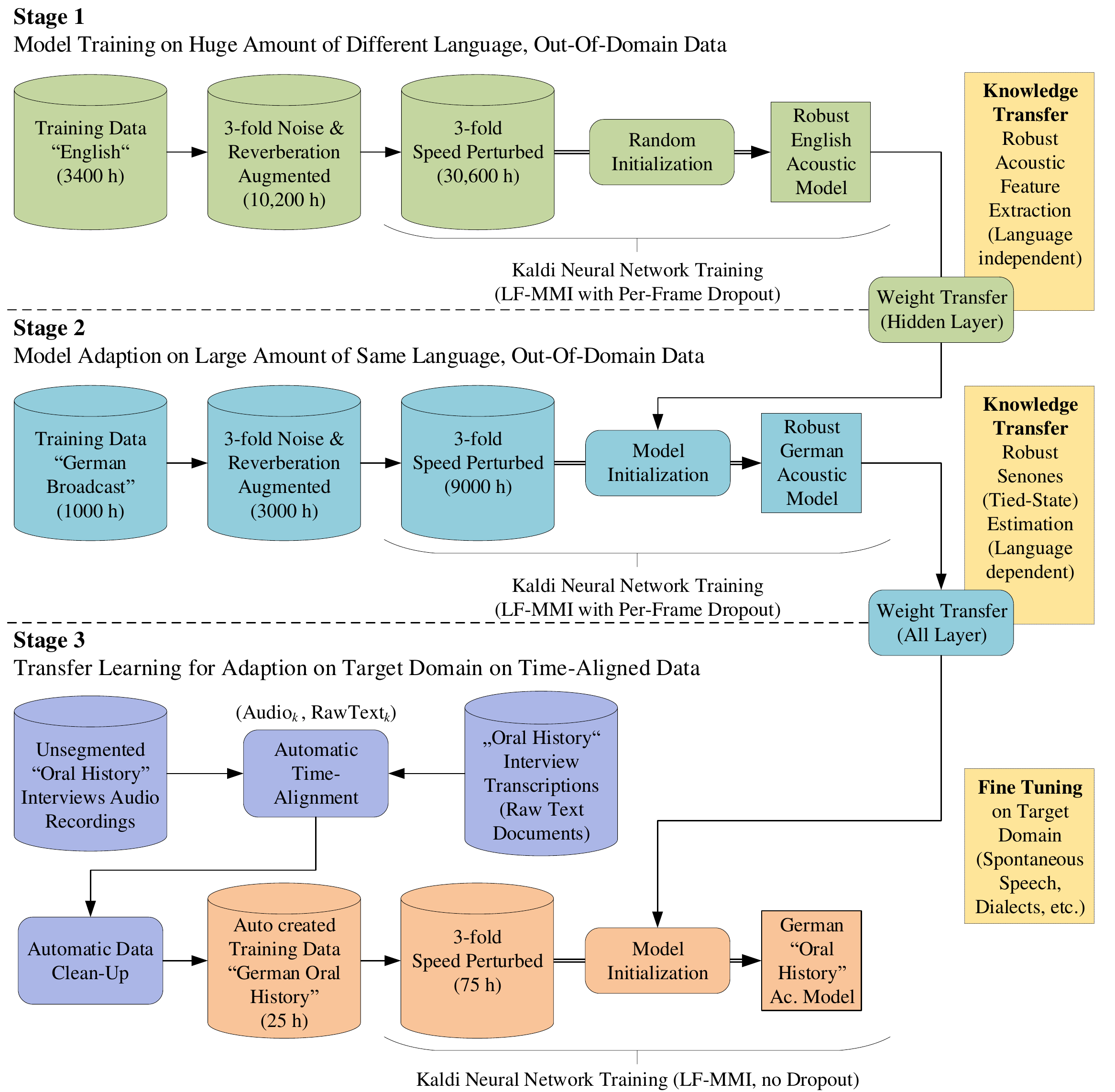} 
	\caption{Proposed approach for cross-lingual, multi-staged acoustic model adaption.}
	\label{fig:multi-staged_domain_adaption}
\end{figure*}

The proposed approach consists of three proceeding stages of acoustic model adaption, as shown in Figure \ref{fig:multi-staged_domain_adaption}. Throughout the stages, we first use a large amount of data from more far-away domains and then smaller amounts of data from nearby domains for training. At the transition of each stage, we transfer the corresponding learned knowledge to the network in the next stage. 

\subsection{Data Augmentation}
To achieve robustness to acoustic conditions of the models trained in the first two stages, we apply noise and reverberation data augmentation increasing the data threefold. Defining discrete-time-signals as sequences of sample values, the applied augmentation can be described as
\begin{equation}
	\label{eqn:data_aug_with_noise}
	({x}_n)_{n \in \mathbb{N} } := (s_n)_{n \in \mathbb{N} } * (h_n)_{n \in \mathbb{N}} + (w_n)_{n \in \mathbb{N} } * (\tilde{h}_n)_{n \in \mathbb{N} }
\end{equation}

if both noise and reverberation inside a simulated room affects the speech signal. Here, $ * $ is the convolution operation for sequences, $ (s_n)_{n \in \mathbb{N} } $ the sequence of the clean speech signal, $ (h_n)_{n \in \mathbb{N} } $, $ (\tilde{h}_n)_{n \in \mathbb{N} } $ are room impulse responses modeling the reverberation of one room at different positions and $ (w_n)_{n \in \mathbb{N} } $ describes the sequence of the noise signal. 

If only reverberation and no background noise affects the speech signal, $ \forall n \in \mathbb{N}: w_n = 0 $ applies and yields 
\begin{equation}
	\label{eqn:data_aug_without_noise}
	(x_n)_{n \in \mathbb{N} } := (s_n)_{n \in \mathbb{N} } * (h_n)_{n \in \mathbb{N}}. 
\end{equation}

\subsection{Stage 1: Training on Huge Amount of English Data}
In the first stage of the proposed approach, as shown in Figure \ref{fig:multi-staged_domain_adaption}, a robust acoustic model is trained using a huge amount of different-language, out-of-domain training data. A model that is well trained on a large amount of data has learned to perform a robust extraction of relevant acoustic input features and learned useful internal representations for the classification task. We assume that these aspects are, at least to some extent, language independent for related languages. To use this knowledge for tasks in languages with less available data, we apply a weight transfer of all hidden layers and use these layers to initialize the acoustic model neural network training in the second stage.

Currently, probably English is the language with the largest amount of available training data for speech recognition. Therefore, we propose combining several English corpora from different domains to train the acoustic model in this stage. Furthermore, in our approach we apply the aforementioned noise and reverberation data augmentation to further increase the robustness and generalization of the model increasing training data three-fold.

As a default step of the acoustic model training, we train an i-vector extractor on the English data in this stage. We further use this i-vector extractor in the following two stages for the acoustic model adaption training. Currently, we do not retrain the extractor on data from the target language.

\subsection{Stage 2: Transfer Learning on German (Out-Of-Domain) Broadcast Data}
In the second stage, the model is adapted to the language of the target domain, however, with training data from another domain. We perform the cross-lingual adaption implicitly by replacing the language-dependent output layer of the neural network trained on English with a randomly initialized output layer for German tied states---and then train this model on German data. No manual phoneme mapping is applied. 

In this stage, the feature extraction and representation learning of the lower layers is further improved for the target language. The classification of subphonetic units, which are language dependent, is learned in the output layer. For our task, we use 1000 hours of German broadcast data in this stage. Again, in this stage training data is increased three-fold using noise and reverberation data augmentation with slightly different configuration than in the previous stage.

The training setup in this stage is completely equal to the one used in the previous stage. Parameter tuning, such as adapting the learning rate layer-wise, might lead to better results. However, due to the large computational load to train a state-of-the-art acoustic model neural network on 1000 hours, we leave this for future work.

\subsection{Stage 3: Transfer Learning on Automatically Aligned German Oral History Interviews}
In the last stage, the acoustic model is finally adapted to the target domain. We do not only transfer the hidden layers in this stage but we apply a full weight transfer of the entire source model for initialization of the target model. In particular, we do not replace the output layer in this stage, since we use the same set of phonemes and the same phonetic decision tree both in Stage 2 and Stage 3. We obtain the respective training lexicons using the same grapheme-to-phoneme (G2P) pronunciation model. 

We apply no dropout in this transfer learning stage. In our experiments, we have not found a dropout setup that has led to a notable improvement using very little training data. The initial learning rate in this stage is 100 times smaller than the final learning rate used for the training in the previous stages. These values are our default learning rates based on our previous experiments. All other training parameters are kept equal in all stages. 

Since, we have lack of training data for German oral history interviews, we apply automatic audio transcript alignment according to \newcite{Manohar.2017.KaldiSegmentationAndASRForArabicMGB3Challenge}, but with neural network based acoustic models trained on German broadcast data, to semi-automatically create in-domain training from several transcribed but not time-aligned oral history recordings.

We consider the transcriptions used to create the data in this stage partly erroneous and incomplete, since they are automatically parsed from several source data to normalized plain text. In our task, these are interview transcriptions by different historians using different styles for annotation, containing remarks and notes in the middle of the text and various document types, such as MS Word and RTF. 

We perform automatic text normalization to write out abbreviations, numbers and dates, and restore casing of words at the beginning of sentences as good as possible. However, we generally perform no manual correction of transcripts or selection of data for alignment. The acoustic model we used to create these alignments is a slightly weaker LF-MMI model from prior works, trained robustly on the 1000 hours of German broadcast data that are also used in Stage 2.

\section{Experimental Setup}
We carry out all experiments using the Kaldi ASR toolkit \cite{Povey.2011.KaldiToolkit}. 
Neural network acoustic models are trained using the aforementioned LF-MMI \cite{Povey.2016.LF_MMI} approach and speed perturbation \cite{Ko.2015.SpeedPerturbation} increasing the amount of data three-fold using constant speed factors $ 0.9 $ and $ 1.1 $.

\subsection{Acoustic Model Neural Network}
\begin{figure}[t]
	\centering
	\includegraphics[width=0.99\columnwidth]{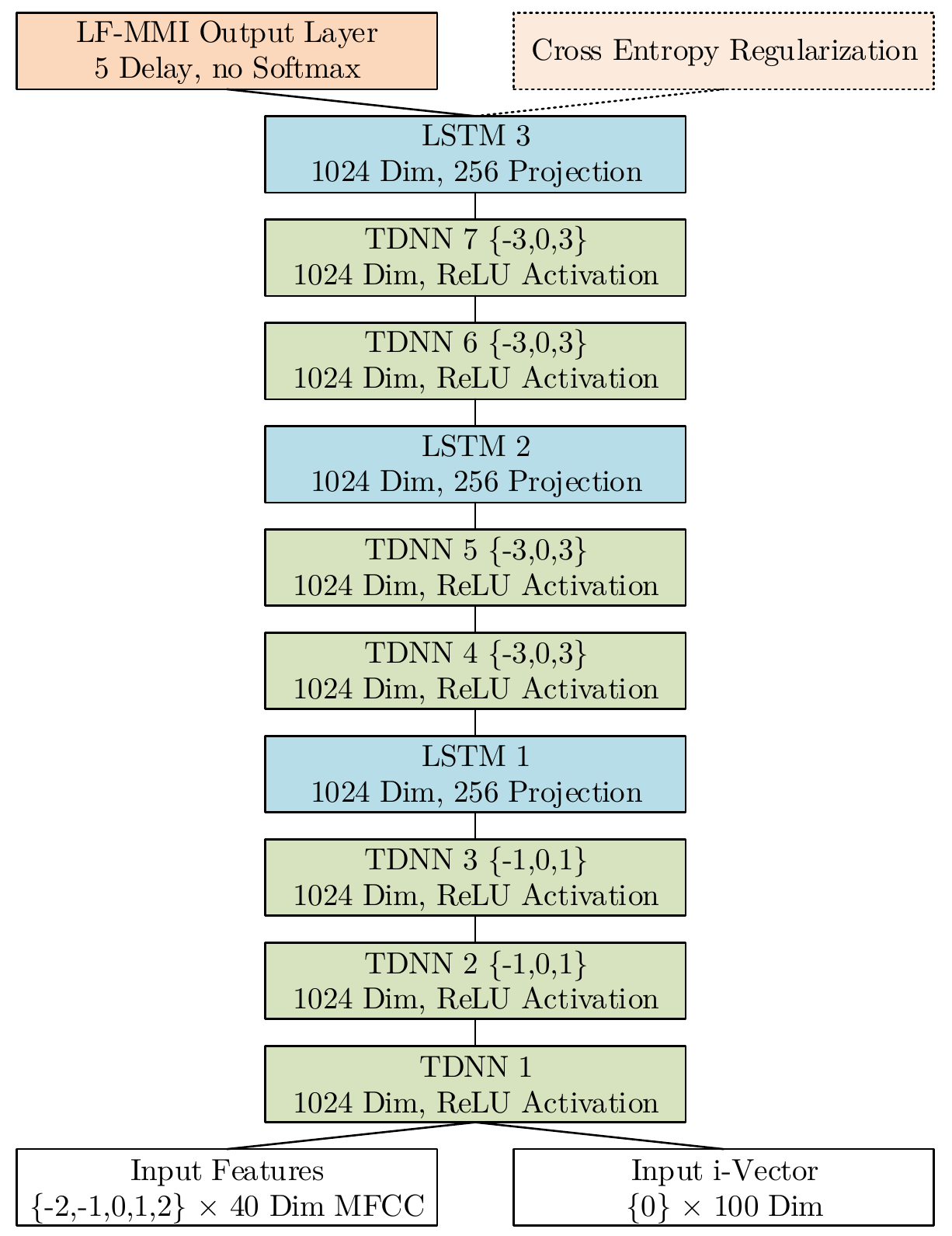} 
	\caption{Architecture of the acoustic model neural network used for our experiments.}
	\label{fig:chain_tdnn_lstm_nnet}
\end{figure}

The acoustic model in our experiments uses a $ 300 $-dimensional input at each time-step consisting of five consecutive $ 40 $-dimensional MFCC features and a $ 100 $-dimensional i-vector \cite{Dehak.2011.ivector}. We use a topology with ten hidden layers that was proposed and investigated by \newcite{Cheng.2017.KaldiLSTMDropout} with seven TDNN layers \cite{Waibel.1989.TDNN}, \cite{Peddinti.2015.KaldiTDNN} and three LSTM layers \cite{Hochreiter.1997.LSTM} stacked in the order given in Figure \ref{fig:chain_tdnn_lstm_nnet}. The applied implementation uses LSTM layers with forget gates \cite{Gers.2000.LSTMLearningToForget}, peephole connections \cite{Gers.2000.LSTMPeepholes} and projection layers \cite{Sak.2014.LSTMProjection}. The LSTM layers have a cell dimension of $ 1024 $ and a projection dimension of $ 256 $. The TDNN layers are $ 1024 $-dimensional as well.

\begin{table*}[t]
	\caption{Statistics of the used evaluation sets}\smallskip
	\centering
	\smallskip\begin{tabular}{l|r|r|r|r}
		\textbf{Evaluation Set} & \textbf{Length} & \textbf{$ \varnothing $Segment  }   & \textbf{$ \varnothing $Words Per } & \textbf{$ \varnothing $Words Per}  \\ 
		& [min.] & \textbf{Length} & \textbf{Segment} & \textbf{Second} \\
		\hline
		{Oral History}  	& 211	& 5.3 s	& 11.6	& 2.2	\\ 	
		Interaction			& 49	& 1.1 s	& 3.8	& 3.4 \\ 	
		Spoken QALD-7 		& 15	& 4.3 s	& 6.9 	& 1.6 \\ 
		Challenging Broadcast		& 105	& 10.6 s 	& 29.3 & 2.8 \\ 	
		DiSCo  (Broadcast Eval)		& 323	& 2.4	s 	& 6.9 	& 2.9 \\	
	\end{tabular}
	\label{tab:eval_sets}
\end{table*} 

\subsection{Training of English Model in Stage 1}
For training the English model in Stage 1, we combined the training data sets from the well-known corpora \textit{Librispeech} \citelanguageresource{Panayotov.2015.librispeechCorpus}, \textit{Commonvoice}\footnote{voice.mozilla.org}, \textit{Switchboard} \citelanguageresource{Godfrey.1992.Switchboard} and \textit{Fisher} \citelanguageresource{Cieri.2004.Fisher}. Overall, the English data used in training comprises more than 3400 hours of annotated speech. Additionally to the clean data set, we create two distorted versions of the data using noise and reverberation data augmentation according to Equation \ref{eqn:data_aug_with_noise}. The first version uses a signal-to-noise ratio between 5 and 10 dB, the second one between 10 to 20 dB. In both versions, we utilize 266 room impulse responses of small and medium-sized rooms for reverberation and several noises recorded in real-life scenarios.

Using a general propose language model for decoding that is based on crawled data, the English model achieves $ 9.17 \% $ word-error-rate on Librispeech test-clean and $ 18.30 \% $ on the data from \textit{Voices Obscured in Complex Environmental Settings} (\textit{VOiCES}) \citelanguageresource{Richey.2018.VOiCESCorpus}.

\subsection{Adaption to German in Stage 2}
To adapted the English model from Stage 1 to German, we utilize the 1000-hour large-scale corpus of German broadcast speech \textit{GerTV1000h} \citelanguageresource{Stadtschnitzer.2014.GerTV1000hCorpus} in this stage. We considered this data out of domain for the oral history task, since broadcast recordings differ from oral history in terms of the used recording technology, audio signal quality and speech characteristics. Broadcast recordings are generally recorded in clean conditions using professional equipment and expertise. In most cases, the speech is clearly pronounced, well articulated and easy to understand. In contrast, oral history interviews are often recorded using conventional recording equipment that was common at the time of recording. The speakers often have a large distance to the microphone and utter sentences spontaneously as they come to their mind. A characteristic attribute of many oral history interviews are the age- and health-related changes in the way elderly people speak.

To perform the noise and reverberation data augmentation in this stage, two additional distorted versions of the clean training are created. In the first version, we apply artificial reverberation according to Equation \ref{eqn:data_aug_without_noise} using 266 room impulse responses of small and medium-sized rooms. The second one is created according to Equation \ref{eqn:data_aug_with_noise} with both reverberation and noises using a random signal-to-noise ratio between 10 and 20 dB.

\subsection{Adaption to Oral History Target Domain in Stage 3}
Randomly selected oral history interviews of 150 contemporary witnesses serve as data from the target domain for the automatic transcript alignment and domain adaption in Stage 3. It is ensured that only interview of speakers that do not appear in the test data are used for training.
 
The recording quality differs from barely understandable to good quality with only light distortions. These are mostly low energy noises from cassette recording machines or reverberation due to a large distance to the recording microphone in a medium-sized room. Some speakers have a slight dialect or accent, but most speak High German.  

Different historians throughout many years transcribed these interviews using varying annotation styles and formats. Since we performed no manual check, we could not guarantee that the entire transcription of the interview is present for alignment. For the final training we use 25 hours of automatically aligned speech.

\begin{table*}[t]
	\caption{Word error rates of the proposed approach compared to two baselines and the ablation study. Results are reported both for the default and the large decoding language model.}\smallskip
	\centering
	\smallskip\begin{tabular}{l|l|r|r|r|r|r|r}
		& \textbf{}	  		& \textbf{Baseline} 	& \textbf{Baseline}  	& \textbf{Removing} & \textbf{Removing} & \textbf{Removing} & 	\textbf{Proposed}  \\
		& \textbf{}  			& \textbf{Broadcast} 	& \textbf{Oral History} & \textbf{Stage 1} & \textbf{Stage 2} & \textbf{Stage 3}	&  \textbf{Approach}  \\  
		\hline
		\multirow{3}{*}{\rotatebox[origin=c]{90}{\textbf{Setup}}} & Stage 1	(English) & 				& 				& 				& 	$ \times $	& 	$ \times $	& 	$ \times $	\\
		& Stage 2 (German Broadcast) & $ \times $	& 				& 	$ \times $	& 				& 	$ \times $	& 	$ \times $	\\		
		& Stage 3 (Ger. Oral History) &				& 	$ \times $	& 	$ \times $	& 	$ \times $	& 				& 	$ \times $	\\		
		\hline  \hline
		\multirow{10}{*}{\rotatebox[origin=c]{90}{\textbf{Evaluation Set}}} & {Oral History}		 & 27.66	& 37.42	& 25.93	& 28.67	& 27.38	& \textbf{25.91}		\\   
		& \noindent\hspace*{3mm} + larger decode LM	& 27.08	& 38.18	& 25.26 & 28.48	& 26.48	& \textbf{25.17}		\\ 
		\cline{2-8}
		& Interaction				& 48.23	& 69.07	& 48.17	& 58.60	& 47.35	& \textbf{47.14} \\  
		& \noindent\hspace*{3mm} + larger decode LM	& 51.22	& 71.96	& 51.36	& 60.13	& 50.58	& \textbf{50.26}		\\ 
		\cline{2-8}
		& Spoken QALD-7			& 18.95	& 36.74	& 19.09	& 31.08	& 18.61	& \textbf{18.40} \\
		& \noindent\hspace*{3mm} + larger decode LM	& 14.79	& 30.06	& 14.25	& 24.61	& 13.77	& \textbf{13.57}		\\ 
		\cline{2-8}
		& Challenging Broadcast	& 19.68	& 33.55	& 19.81	& 26.58	& 19.47	& \textbf{19.38} \\  
		& \noindent\hspace*{3mm} + larger decode LM	& 17.42	& 31.90	& 17.61	& 24.59	& \textbf{17.34} & 17.36	\\ 
		\cline{2-8}
		& DiSCo (Broadcast Eval)	& 11.89	& 27.91	& 12.45	& 20.81	& \textbf{11.85} & 12.37 \\   
		& \noindent\hspace*{3mm} + larger decode LM	& 12.15	& 28.73	& 12.48	& 20.75	& \textbf{12.12} & 12.42	\\ 
	\end{tabular}
	\label{tab:results}
\end{table*}

\subsection{Evaluation Sets}

We do not only report results on the aforementioned oral history test set but overall on four additional German in-house test sets from different domains. Some of these sets partly share some challenges of oral history interviews, such as spontaneous speech with unclear pronunciations, noises and reverberation. The sets used in our work are listed in Table \ref{tab:eval_sets} along with some statistics characterizing the sets. Along with the overall test set length, the average segment length, the average amount of words per segment, and the average spoken words per second are reported. Very short word sequences per segment usually tend to benefit less from a large powerful language than segments with rather long word sequences. A high number of spoken words per second in a test set indicates a fast speaking speed. The faster speakers are speaking, the more difficult it gets for the speech recognition system---as it gets for human listeners as well---to distinguish different phonemes and words.

As evaluation set for the target domain, we use the German oral history data set from \cite{Gref.2018.oralhistory01}. It consists of $ 3.5 $ hour audio from $ 35 $ different speakers recorded in real oral history interviews. 
The recordings took place between 1980 and 2012, representing a wide range of recording technology, interview methodology, dialects and pronunciations. The set is manually transcribed and segmented. 

The \textit{Interaction} evaluation set contains recordings of people informally talking to each other recorded in challenging acoustic conditions. This set is characterized by very fast, partly overlapping, highly colloquial speech, speaker noises such as laughter and unclear pronunciation. 
This set has the highest average speed of speaking of all used test sets. Furthermore, due to the nature of fast, informal conversations, the average segment length is the shortest and many segments only contain one or two words.
This makes this test set one of the most challenging evaluations sets in our data collection.

The \textit{Spoken QALD-7} corpus contains in-house recorded questions for a question answering system described in \cite{Usbeck.2017.promptsForSpeechAssistantEvalSet}, where several speakers used a web interface and their respective microphone---a headset or build-in laptop microphone for instance. One segment usually comprises exactly one question prompt.

The \textit{Challenging Broadcast} evaluation set contains a $ 1.75 $ hour collection of rather challenging interviews and recordings from the German broadcast domain with a lot of spontaneous speech, often in challenging acoustic conditions, including overlapping and dialectal speech. The average speaking speed is somewhat between the oral history and the interaction test set.

\textit{DiSCo} \citelanguageresource{Baum.2010.DiSCoCorpus} is a corpus for the German broadcast domain and contains four evaluation sets: \textit{planned} and \textit{spontaneous} speech, each in \textit{clean} and \textit{mixed} acoustic conditions. For the sake of clarity, we always report the unweighted mean value of these subsets.

\subsection{Lexicon and Language Model}
\label{subsec:g2pAndLM}
The training and decoding lexicons are obtained using a G2P model trained with Sequitur G2P \cite{Bisani.2008.SequiturG2P}. This model is created using the German pronunciation dictionary \textit{Phonolex} from the \textit{Bavarian Archive for Speech Signals}. 

To assure the improvements of the acoustic model adaption using our proposed approach are consistent and do not depend on a certain language model, we perform two independent decodings in each experiment using two different language models. First, we use our default 500,000 words broadcast language model. This model is trained on broadcast text corpora consisting of $ 75 $ million words. Further, we use a larger broadcast language model that is trained on texts with $ 1.6 $ billion running words of crawled German news data websites. The lexicon of this model contains more than $ 2 $ million different words. Both models are trained as 5-gram language models with a 1e-8  pruning factor. All decoding parameters are kept to fix default values for all of our experiments. Especially, the language model weight is kept to a fixed value for all experiments.

\subsection{Performed Experiments}
We use two baseline models to compare our proposed approach. These models are trained without neural network initialization from a prior model and not further adaption is performed. As the first baseline serves a model trained from scratch on the 1000 hours of German broadcast data with the same configuration and setup as in Stage 2. The second baseline is a model that we train from scratch on the 25 hours of automatically aligned oral history data. Furthermore, to determine which of the stages contributes to the improvements to what extent, we carry out ablation study experiments in which we omit one of the stages from training. 

In each experiment in which the English training in Stage 1 is skipped (both in the baselines and in the ablation study), the i-vector extractor is also trained with the corresponding data used in the first respective training step. To better understand, where the increased robustness through knowledge transfer from Stage 1 comes from, we also perform some experiments comparing the German and English trained i-vector extractor.

\section{Results}
The results of the proposed approach, baseline models, and the ablation study are summarized in Table \ref{tab:results}. For each test set, the upper row shows the achieved word error rate using the default 500,000 words language model. The respective next row shows the results with the larger 2-million-words language model.

\subsection{Comparison to Baselines}
The oral history baseline performs significantly worse than the broadcast baseline on all sets---even on the target domain. This is due to the significantly smaller amount of training data. Compared to the broadcast baseline, the proposed approach achieves a relative word error improvement of $ 6.3 \% $ on the oral history test set using the default language model. Using the larger language model, the relative improvement is $ 7.1 \% $.

On all other evaluation sets, except DiSCo, we observe an improvement using the proposed approach. This indicates a good generalization of the model and suggests that the model is robust enough for operation in real-world applications. The slightly reduced performance on the DiSCo set is probably because this test data is already very close to the conditions presented by the broadcast training used to train the baseline model.

\subsection{Ablation Study}

\subsubsection{Removing Stage 1}
In the first ablation study setup, we remove the adaption from the English model trained in Stage 1. In particular, we randomly initialize the model in Stage 2 for training on German broadcast data and then adapted to the oral history data in Stage 3. Furthermore, we train a new i-vector extractor on German data. Compared to the broadcast baseline, we observe a significant word error rate improvement using both language models on the oral history test set. The achieved word error rate on this test set is almost as good as offered by the proposed approach. However, compared to the broadcast baseline, on the Challenging Broadcast and DiSCo test set, the results get worse with both language models in this setup. This is also true for the QALD-7 test set decoded with the smaller language model and for the interaction test set and the larger language model. This indicates that the initialization with the English trained model increases the robustness of the proposed approach for many domains.

\subsubsection{Removing Stage 2}
In this setup, we skip the adaption to German broadcast in Stage 2 and instead directly adapt the English trained model both to the target language and target domain using the oral history training data. While the results are still worse than the broadcast baseline, a remarkable improvement is achieved compared to the oral history baseline. Using only English training data and 25 hours of automatically aligned German training data, we achieve a word error rate reduction from $ 37.4 $ to $ 28.7 \% $ on the target domain decoding with the default language model. Results are similar for the large language model. 

This value is only one absolute percentage point higher than the value achieved by the broadcast baseline that we trained on 1000 hours of manually annotated German data. Thus, adapting from a rich-resourced language directly to the target language and domain is a reasonable approach if no other data is available for training in the target language. In particular, this is shown in this experiment for state-of-the-art LF-MMI acoustic models with an LSTM-TDNN topology, which usually require a lot of data during training. For real-world applications one cannot expect the model to be as robust as models trained robustly on large-scale data from the target language---but definitely more robust than training from scratch on the small data set only which can be usefully especially for under-resourced languages. 

\subsubsection{Removing Stage 3}
In the last setup of the ablation study, we omit the fine-tuning to target domain and evaluate the model trained on broadcast data in Stage 2. Compared to the broadcast baseline, this experiment shows once again that initialization with the English-trained model leads to an increase in robustness and better results on all test sets. Remarkably, this model even achieves the best results on the DiSCo test set which we consider close to the German broadcast training data. This setup also achieves slightly better results than the proposed approach on the Challenging Broadcast test set with the larger language model. However, the gain is less than $ 0.2 \% $ relative.

\subsection{Influence of the i-Vector Extractor}
In the following experiments, we study whether the improved robustness through knowledge transfer of the English model is due to weight transfer, as we expect, or due to better i-vector extractor trained on English data instead of German. Therefore, we compare each combination of using German or English trained i-vector extractors each with a random model initialization or initialization of hidden layers with the English trained model when training on the 1000 hour German broadcast data. For simplicity, we only report results on the small language model. The results are summarized in Table \ref{tab:results_ivector}.

\begin{table}[t]
	\caption{Comparison of German and English trained i-vector extractors in acoustic model training on German broadcast data with random model initialization and with knowledge transfer from English-trained model.}\smallskip
	\centering

	\smallskip\begin{tabular}{l|l|r|r|r|r}

		\multirow{3}{*}{\rotatebox[origin=c]{90}{\textbf{Setup}}} & & & & & \\
		& {Ac. Model Init.} 	& Rand. 	& Eng.	& Rand.	& Eng. \\
		& {i-vector}		& Ger.	& Ger.	& Eng.	& Eng. \\
		\hline
		\hline
		\multirow{5}{*}{\rotatebox[origin=c]{90}{\textbf{Eval Set}}} & {Oral History}  & 27.66		& 27.75	& 27.70	& \textbf{27.38}	\\ 
		& Interaction		& 48.23		& 48.54	& 47.83	& \textbf{47.35}	\\ 
		& Spok. QALD-7 	& 18.95		& 19.15	& 19.43	& \textbf{18.61}	\\ 
		& Chall. Broadc.	& 19.68 	& 19.78 & 19.57	& \textbf{19.47}	\\ 	
		& DiSCo 			& 11.89		& 11.86	& 11.88	& \textbf{11.85}	\\	

	\end{tabular}

	\label{tab:results_ivector}
\end{table}

Using the English-trained i-vector extractor along with acoustic model initialization from the English-trained model---as is the case in our proposed approach---we achieve the best results on all test sets. Using the German i-vector extractor with English model initialization leads to the worst result for three out of the five test sets. This is to be expected since i-vectors of similar speakers from two differently trained i-vector extractors point to completely different directions in the two different 100-dimensional vector spaces. Since the English model is trained with i-vectors from one space, using different i-vectors in the second stage causes wrong estimations of speakers and this relation has to be relearned for the new vector space in Stage 2.

Using the English i-vector extractor instead of the German one with random acoustic model initialization only leads to slightly better results on three test sets and slightly worse results on two test sets. Therefore, we conclude that, in deed, weight transfer leads to better results---and not a better i-vector extractor trained on English data. However, in order to use weight transfer sensibly, the i-vector extractor must be used, which was also used to train the original model.

\section{Conclusions}
In this work, we proposed and investigated an approach that performs a robust acoustic model adaption to a target domain in a cross-lingual, multi-staged manner. Our approach addresses challenges where only little training data for the target domain are present and enables the exploitation of large-scale training data from other domains in both the same and other languages. These other domains may be acoustic distortions such as reverberation or poor recording equipment, but also deviations in the way of speaking, articulation, dialects, spontaneous speech, unclear articulations and much more.

We studied our approach using the challenging example of German oral history interviews with the aim to obtain an acoustic model that is robust enough to be applied in real-world applications. In our experiments we first trained a robust acoustic model for English with more than 3000 hours of data, that we then adapted to German using 1000 hours of German broadcast data. These model is again adapted using 25 hours of in-domain oral history interview.

To determine the robustness of the model, for this case study we performed exhausting experiments and evaluated not only on in-domain data but also on several German test sets from other domains and two different decoding language models. To thoroughly determine which stage of the proposed approach leads to which improvement, we further performed different ablation study experiments. Thereby we have shown that the direct adaptation of LF-MMI acoustic models from one language to another leads to good results, even using only very little training data from the target domain. Thus, this observation can contribute to the ongoing research on speech recognition for under-resourced language.

The model trained with our proposed approach achieves a relative reduction of the word error rate by more than 30\% compared to a model trained from scratch only on the target domain, and 6--7\% relative compared to a model trained robustly on 1000 hours German broadcast training data.

\section{Acknowledgements}
This research has been funded by the Federal Ministry of Education and Research of Germany (BMBF) in the project \textit{KA$^3 $ - K{\"o}lner Zentrum f{\"u}r Analyse und Archivierung von AV-Daten} (Cologne center for the analysis and archiving of audiovisual data) (project number: 01UG1811B).

\section{Bibliographical References}\label{reference}

\bibliographystyle{lrec}
\bibliography{literature}

\section{Language Resource References}
\label{lr:ref}
\bibliographystylelanguageresource{lrec}
\bibliographylanguageresource{languageresource}

\end{document}